\documentclass[a4paper,fleqn]{cas-dc}

\usepackage[sort&compress,numbers]{natbib}
\usepackage[utf8]{inputenc}
\usepackage{textcomp}
\DeclareUnicodeCharacter{22EF}{\dots}
\DeclareUnicodeCharacter{0302}{\^}
\DeclareUnicodeCharacter{0301}{\'}
\usepackage{soul}

\def\tsc#1{\csdef{#1}{\textsc{\lowercase{#1}}\xspace}}
\tsc{WGM}
\tsc{QE}
\tsc{EP}
\tsc{PMS}
\tsc{BEC}
\tsc{DE}

\begin{document}
\let\WriteBookmarks\relax
\def\floatpagepagefraction{1}
\def\textpagefraction{.001}
\shorttitle{Dewar-Anthracyne}
\shortauthors{Laranjeira et~al.}

\title [mode = title]{A Novel Graphyne-Like Carbon Allotrope: 2D Dewar-Anthracyne}

\author[1]{José A. S. Laranjeira}
\affiliation[1]{
organization={Modeling and Molecular Simulation Group},
addressline={São Paulo State University (UNESP), School of Sciences}, 
city={Bauru},
postcode={17033-360}, 
state={SP},
country={Brazil}}

\credit{Conceptualization of this study, Methodology, Review and editing, Investigation, Formal analysis, Writing -- review \& editing, Writing -- original draft}
\author[2]{Kleuton A. L. Lima}
\affiliation[2]{
organization={Department of Applied Physics and Center for Computational Engineering and Sciences},
addressline={State University of Campinas}, 
city={Campinas},
postcode={13083-859}, 
state={SP},
country={Brazil}}
\credit{Conceptualization of this study, Methodology, Review and editing, Investigation, Formal analysis, Writing -- review \& editing, Writing -- original draft}
\author[1]{Nicolas F. Martins}
\credit{Conceptualization of this study, Methodology, Review and editing, Investigation, Formal analysis, Writing -- review \& editing, Writing -- original draft}
\author[3]{Luiz A. Ribeiro Junior}
\affiliation[3]{
organization={Computational Materials Laboratory, LCCMat, Institute of Physics},
addressline={University of Brasília}, 
city={Brasília },
postcode={70910‑900}, 
state={DF},
country={Brazil}}
\credit{Supervision, Funding Acquisition, Review and editing, Formal analysis, Writing -- review \& editing, Writing -- original draft}

\author[2]{Douglas S. Galvao}
\credit{Conceptualization of this study, Methodology, Review and editing, Investigation, Formal analysis, Writing -- review \& editing, Writing -- original draft}

\author[1]{Julio R. Sambrano}
\cormark[1]
\cortext[cor1]{Corresponding author}
\credit{Conceptualization of this study, Methodology, Review and editing, Investigation, Formal analysis, Writing -- review \& editing, Writing -- original draft}

\begin{abstract}
Anthracyne (2DDA). 2DDA consists of chains of Dewar-anthracenes connected by acetylenic linkages. DFT-based simulations show that 2DDA is thermally stable and exhibits no imaginary phonon modes, confirming its dynamic stability. 2DDA is metallic with Dirac-like features near the Fermi level, dominated by C p$_z$ orbitals. It shows marked mechanical anisotropy, with Young's modulus of  176.24 N/m (x) and 31.51 N/m (y), shear modulus up to 69.14 N/m, and Poisson’s ratio varying from 0.27 to 0.87. The material also exhibits strong anisotropic optical absorption in the visible and ultraviolet ranges. Raman and IR spectra reveal intense bands at 648 cm$^{-1}$ (Raman) and 1292 cm$^{-1}$ (Infrared). Nanoribbon structures derived from 2DDA exhibit diverse electronic behaviors, from metals up to bandgap values of up to 0.42 eV, depending on the edge-type terminations and width. These findings demonstrate the 2DDA potential for nanoelectronic and optoelectronic applications.
\end{abstract}

% \includegraphics{figs/grabs.pdf}

%\item 2D Dewar-Anthracene is a novel porous carbon network designed via DFT.
%\item 2D Dewar-Anthracyne is a novel porous 2D carbon allotrope.
%\item The structure displays Dirac-like metallic behavior.
%\item It consists of anthracene-based chains connected by Dewar-type rings.
%\item It possesses strong mechanical anisotropy with Young's modulus of up to 176.24 N/m, which enables energy storage applications.
%\item Pronounced optical absorption in the visible and UV ranges supports applications in optoelectronics.


\maketitle

\section{Introduction}

In recent years, nanotechnology research has made significant progress, both theoretically and experimentally. One of its highlights is the exploration of two-dimensional materials, driven by the discovery of graphene \cite{geim2007rise} and its remarkable mechanical, electronic and optical properties\cite{zhuang2015two}, which include exceptional structural strength, high carrier mobility, and chemical stability \cite{choi2010synthesis}.

%However, its zero band gap and nonporous structure limit its applications in digital electronics and optoelectronics \cite{huang2020chemistry,nguyen2016promising}, motivating the search for alternative 2D carbon allotropes with tunable properties.

%Among the most promising strategies is incorporating non-hexagonal rings and structural motifs derived from well-known organic molecules \cite{luo2024non,obermann2024wavy,fernandez2022synthetic}.

The latest advances in computational materials science have enabled the theoretical design of novel carbon-based two-dimensional networks with diverse topologies, tunable porosity, and electronic properties ranging from metallic to wide-bandgap semiconductors, surpassing the limitations of graphene \cite{curtarolo2013high, jain2013apl, tiwari2016magical, hirsch2010era, enyashin2011graphene}. Among them, porous 2D carbon materials are attractive due to their high surface area, accessible channels, and intrinsic chemical tunability \cite{zheng2015two}. By providing controlled pore size and distribution, these materials can perform excellently in applications such as membranes, gas sensing, ion transport, and energy storage\cite{huang2020chemistry, nawaz2025flatland}. Significantly, porosity also affects the material's electronic band structure and charge distribution, often enabling direction-dependent conductivity, band gap opening, and adsorption selectivity \cite{liu2017porous}. Recently, new carbon-based monolayers with attractive topologies, such as irida-graphene \cite{junior2023irida}, THD-C \cite{jenkins2024thd}, TH-graphyne \cite{lima2025th}, naphthyne \cite{laranjeira20252d}, anthraphenylene \cite{lima2025anthraphenylenes}, and many others \cite{li2025promising, cavalheiro2024can, katin2024diamanes, ullah2024theoretical} have been proposed.

In parallel, 1D nanoribbons obtained from 2D materials provide additional possibilities for property tuning, particularly via edge engineering and quantum confinement effects \cite{wang2021graphene}. Graphene nanoribbons (GNRs), for example, can exhibit semiconducting or metallic behavior depending on their width and edge orientation (zigzag or armchair) \cite{son2006energy,han2007energy}. This versatility extends to other 2D allotropes, whose nanoribbon forms can present unique bandgap values, magnetic edge states, or topological features not present in their 2D form \cite{dutta2010novel}. These characteristics make nanoribbons highly attractive for nanoelectronic, spintronic, and thermoelectric applications \cite{celis2016graphene}.

%Despite recent advances in synthesizing novel 2D carbon allotropes \cite{fan2021biphenylene,hou2022synthesis,desyatkin2022scalable,aliev2025planar,toh2020synthesis,tian2023disorder,meirzadeh2023few,pan2023long,bai2024nitrogen,li2010architecture,liu2022constructing}, other carbon-based nanomaterials that simultaneously offer intrinsic porosity, mechanical and optical anisotropy, and Dirac-like electronic behavior while maintaining structural robustness and thermal stability are still needed. Designing such materials is essential to guide potential synthesis routes and to enable the development of more efficient carbon-based nanostructures. This trend continuously motivates the proposal of a new carbon architecture that integrates all these features in a single framework.

However, creating viable synthetic routes for these monolayers remains a significant challenge. Recent progress in bottom-up fabrication techniques has demonstrated the potential for experimentally realizing some predicted 2D carbon allotropes. A crucial factor in these methods is the strategic selection of the molecular precursors combined with precise control over thermodynamic and kinetic parameters to achieve nanoscale structural accuracy. For instance, the synthesis of 2D biphenylene involved the self-assembly of poly(2,5-difluoro-para-phenylene) (PFPP) on an Au(111) substrate \cite{hudspeth2010electronic}. The resulting fused 4–6–8 carbon ring system, initially predicted from theoretical studies, was experimentally verified as metallic. Likewise, graphenylene, another material derived from biphenylene, was successfully synthesized via polymerization of 1,3,5-trihydroxybenzene, yielding a dodecagonal ring with a 5.8 \r{A} diameter \cite{du2017new}, in close agreement with computational predictions \cite{fabris2018theoretical}.

In this framework, Dewar-anthracene - a metastable anthracene isomer with a strained bicyclic structure - presents a promising molecular structural unit for creating innovative 2D carbon networks \cite{PRITSCHINS19821151}. Its synthesis follows a multistep strategy, where a benzocyclobutadiene unit undergoes a Diels–Alder reaction with an activated dienophile, such as 3,6-dihydrophthalic anhydride, followed by oxidative bis-decarboxylation and controlled dehydrogenation processes \cite{iwata2021synthesis, applequist1964synthesis}. This approach enables the reversible modulation of aromaticity, facilitating the formation of extended conjugated structures.

One important class of carbon allotropes is graphynes (GYs) and graphdiynes (GDYs). These materials, like graphene, form atomically thin, planar networks but differ in structure due to the presence of acetylenic (–C$\equiv$C–) or diacetylenic (–C$\equiv$C–C$\equiv$C–) groups. This structural modification results in a hybrid structure composed of sp$^2$-hybridized carbon atoms (three-fold coordinated) with sp$^1$ ones (two-fold coordinated),  topologically between graphene (entirely sp$^2$) and carbyne (entirely sp$^1$) \cite{huang2018progress}. 

Due to their extended $\pi$-conjugation network, porous architecture, and lower density than graphene, these sp$^2$+sp$^1$ networks exhibit tunable electronic properties, which make them promising candidates for nanoelectronics, membrane technologies (such as hydrogen separation) \cite{yeo2019multiscale, yeo2019multiscale, apriliyanto2018nanostructure}, energy storage \cite{li2011high, liu2014hydrogen, gangan2019first}, and battery anode materials \cite{mao2020graphyne, yang2020nitrogen, shomali2019graphyne}. 

In this work, a new 2D carbon allotrope combining the structure of Dewar-antracene and graphynes: 2D Dewar-Anthracyne (2DDA). 2DDA consists of chains of Dewar-anthracenes connected by acetylenic linkages \cite{yousef2013excited,applequist1964synthesis} (Figure \ref{fig:system}(a)) was theoretically proposed. Using density functional theory (DFT)-based simulations, it was demonstrated that 2DDA is thermally and dynamically stable, with metallic characteristics, anisotropic mechanical behavior, and strong optical activity in the visible and ultraviolet regions. We have also investigated nanoribbon configurations derived from the 2DDA lattice, obtaining semiconducting behavior for some selected edge geometries.

\section{Methodology}

All DFT simulations were carried out using the Vienna Ab initio Simulation Package (VASP) \cite{kresse1993ab,kresse1996efficient} and CRYSTAL17 \cite{dovesi2005crystal} codes. 

For the VASP simulations, the projector augmented wave (PAW) method was used to describe the interaction between ions and electrons \cite{PhysRevB.50.17953}. The Perdew–Burke–Ernzerhof (PBE) exchange-correlation functional, within the generalized gradient approximation (GGA), was adopted. \cite{PhysRevLett.77.3865}. A plane-wave energy cutoff of 500 eV was employed for all calculations, and the used convergence criteria for the electronic and ionic relaxations were 10$^{-5}$ eV and 10$^{-1}$ eV/\r{A}, respectively. Structural optimization and static electronic calculations were performed using a Monkhorst-Pack k-point mesh of $5\times 5\times 1$. A vacuum layer of at least 20 \r{A} was added along the z-direction to eliminate spurious interactions between periodic/mirror images. For the nanoribbon models, periodic boundary conditions were applied along the ribbon axis, with at least 15 \r{A} for the vacuum buffer layer along the lateral directions. The corresponding Brillouin zones were sampled with a $1\times 5\times 1$ k-point mesh.

To assess the structural thermal stability, ab initio molecular dynamics (AIMD) simulations were performed within the NVT ensemble using a Nos\'e-Hoover thermostat \cite{hoover1985canonical}, at 300 K for a total simulation time of 5 ps, with a time steps of 0.5 fs.

%Elastic constants were computed via the stress–strain method using small finite deformations, from which the Young’s modulus, shear modulus, and Poisson’s ratio were extracted as directional functions in the 2D plane \cite{wu2005systematic,lee2008measurement}. The optical properties, including the real and imaginary parts of the dielectric function, were computed within the random phase approximation (RPA), considering local field effects \cite{gajdovs2006linear}.

To analyze the 2DDA vibrational properties, the coupled perturbed HF/Kohn–Sham algorithm \cite{ferrero2008coupled}, as implemented in the computational package CRYSTAL17 \cite{dovesi2005crystal} was used. The CRYSTAL17 calculations were carried out using a triple-zeta valence with polarization (TZVP) basis set \cite{vilela2019bsse} together with the PBE functional. The structure was optimized by monitoring the root mean square (RMS) and the absolute value of the largest component of both the gradients and the estimated displacements. The adopted convergence criteria in the optimization for RMS and the largest component for gradient were 0.00030 and 0.00045 a.u., and for displacements, 0.00120 and 0.00180 a.u., respectively. The reciprocal space was sampled using Pack-Monkhost and Gilat grids with sublattice and a shrinking factor of 12.

\section{Results and Discussion}
\subsection{2D Dewar-Antracyne}
\begin{figure*}[!htb]
    \centering
    \includegraphics[width=\linewidth]{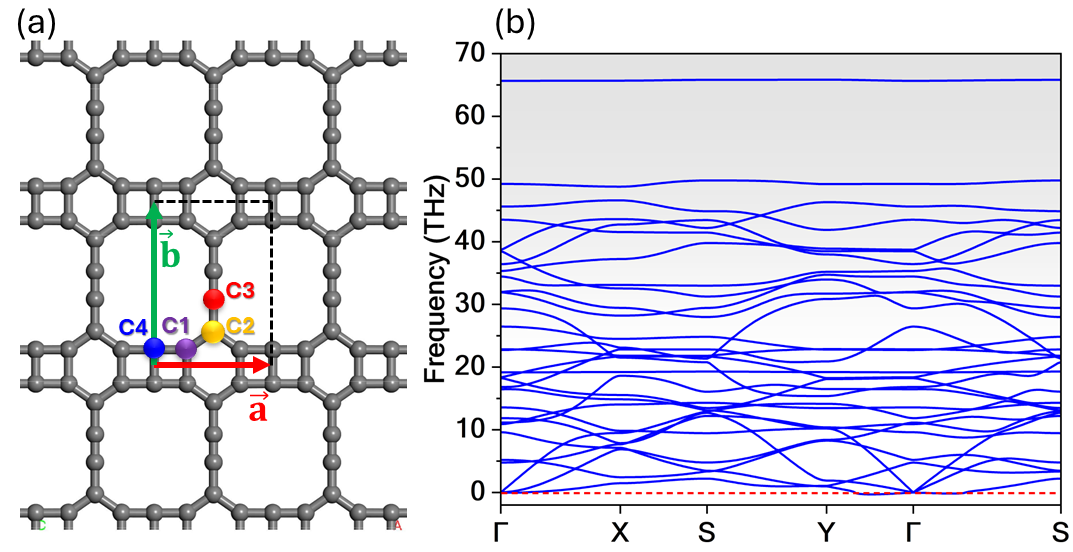}
    \caption{(a) 2D Dewar-Anthracyne (2DDA) top view showing the rectangular unit cell and lattice vectors. (b) Phonon dispersion curves without imaginary frequencies suggest dynamical structural stability.}
    \label{fig:system}
\end{figure*}

The 2DDA atomic structure and phonon dispersion are shown in Figure \ref{fig:system}. The 2DDA periodic framework consists of Dewar-anthracene-like chains aligned along one direction, interconnected by acetylenic groups along the perpendicular axis. The structure has a rectangular symmetry that belongs to the $Pmmm$ (no. 47) space group, with lattice vectors $\vec{a}=5.18$ \r{A} and $\vec{b}=7.14$ \r{A} indicated in red and green colors, respectively. The unit cell consists of four non-equivalent carbon atoms generating 4-, 6-, and 14-membered rings. These atoms are positioned at C1(0.276, 0.105, 0.000), C2(0.500, 0.219, 0.000), C3(0.500, 0.414, 0.000), and C4(0.000, 0.103, 0.000). This arrangement contributes significantly to mechanical anisotropy and electronic delocalization, as discussed below. The dewar-benzene motifs clearly deviate from the ideal $sp^{2}$ configuration, generating a torsional strain in the lattice.

%Moreover, the periodic arrangement of fused six-membered rings and four-membered Dewar rings leads to the formation of uniform pores, suggesting potential for transport and adsorption-related applications.

The cohesive energy (\( E_{\text{coh}} \)) of 2D Dewar-Anthracyne was calculated as \( E_{\text{coh}} = (E_{\text{DA}} - n E_{\text{C}})/n \), where \( E_{\text{DA}} \) is the total energy of 2D Dewar-Anthracyne, \( E_{\text{C}} \) is the energy per carbon atom, and \( n \) is the number of carbon atoms in the unit cell. The resulting formation energy is -7.13 eV/atom, which suggests structural stability. This value is comparable to other predicted carbon allotropes such as graphene (-7.68 eV/atom), T-graphene (-7.45 eV/atom), Graphenylene (-7.33 eV/atom), Graphenyldiene (-6.92 eV/atom), and Graphyne (-7.20 eV/atom), all calculated in the present study. 2DDA is more stable than the Graphenyldiene, which is also a dewar-anthracene-like and a purely $sp^2$ structure. On the other hand, 2DDA has a \( E_{\text{coh}} \) lower than the observed for Graphyne, which can be explained by the torsional tension introduced by the dewar-benzene rings. The 2DDA \( E_{\text{coh}} \) value, combined with the fact that its structural units have already been synthesized, suggests that its experimental realization may be feasible through bottom-up approaches, such as molecular precursor assembly.

%phagraphene (-9.03 eV/atom)~\cite{ghosh2022intriguing}, and higher than \(\alpha\),\(\beta\),\(\gamma\)-graphynes (ranging between 0.6-1.0 eV/atom)~\cite{zhao2013two}, already-synthesized graphdiyne (0.77 eV/atom)~\cite{zhao2013two}. The formation energy of 2D Dewar-Anthracene, combined with its simple composition (pure carbon) and sp\textsuperscript{2}-like bonding, suggests that its experimental realization may be feasible through bottom-up approaches, such as molecular precursor assembly.

Figure \ref{fig:system}(b) shows the phonon dispersion curves calculated along high-symmetry directions of the Brillouin zone. All phonon modes display real and positive frequencies, with no imaginary modes throughout the Brillouin zone. This trend suggests the 2DDA dynamical structural stability. The acoustic branches are well-separated from the optical modes, and the highest phonon frequency reaches approximately 50 THz, indicating strong carbon-carbon bonding typical of s$p^{2}$-hybridized networks.

%When compared to other carbon allotropes, such as graphene (with a maximum phonon frequency near 48–50 THz) \cite{anees2015temperature,diery2018nature} and graphdiyne (15-17 THz, which typically shows lower frequencies due to acetylenic linkages) \cite{zhu2022phononic}, 2D Dewar-Anthracene exhibits a comparable vibrational spectrum in the high-frequency region. This behavior supports the conclusion that the bonding environment in 2D Dewar-Anthracene is similarly robust despite its topological complexity and porosity.

When compared to other carbon allotropes, such as graphene (with a maximum phonon frequency near 48–50 THz) \cite{anees2015temperature,diery2018nature}, 2DDA exhibits a distinct vibrational spectrum, featuring a phononic band gap between 50 THz and 65 THz and a nearly flat band emerging at 65 THz. An analog phenomenon was reported for graphdiyne, where a similar band gap was observed, and flat bands emerged at approximately 65 THz \cite{zhu2022phononic}. This behavior can be attributed to the presence of triple bonds (C$\equiv$C), which have large force constants, leading to high-energy stretching modes. The existence of a phononic band gap in this range suggests that vibrational interactions are significantly influenced by the network connectivity, potentially reducing phonon dispersion in this region. Additionally, the flat band at 65 THz may be associated with localized modes involving the stretching of acetylenic bonds, reflecting the weak coupling of these modes with the rest of the structure.

\begin{figure}[!htb]
    \centering
    \includegraphics[width=\linewidth]{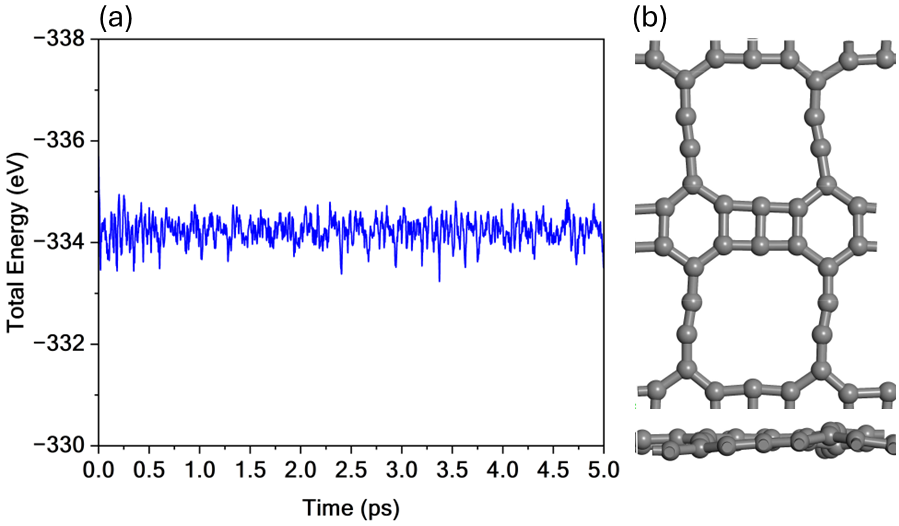}
    \caption{(a) Total energy variation during a 5 ps AIMD simulation at 300 K. (b) Top view of the final structure showing thermal robustness and preserved pore topology.}
    \label{fig:aimd}
\end{figure}

To further assess the 2DDA finite-temperature structural stability, AIMD simulations were performed at 300 K in the canonical (NVT) ensemble. Figure \ref{fig:aimd} summarizes the results of this analysis. 

Figure \ref{fig:aimd}(a) shows the time evolution of the total energy over a simulation run of 5 ps, with time steps of 1 fs. The energy fluctuates around -334 eV with a stable amplitude of less than 2 eV, showing no signs of systematic drift or sudden energy drops that would indicate structural degradation, bond breaking, or significant rearrangements. This energy profile confirms that the structure remains thermodynamically stable under elevated thermal excitation.

The 2DDA final AIMD snapshot is shown in Figure \ref{fig:aimd}(b). The top view shows that the framework's overall structural integrity is preserved, with only minor local distortions, particularly near the Dewar-type rings. These distortions are typical of thermal vibrations and do not affect the integrity of the overall 2D lattice. Importantly, no ring opening or collapse indicates a high kinetic barrier to thermal decomposition.

%Compared to other porous carbon allotropes, such as Holey-Graphyne and TH-graphyne, which may buckle under similar conditions \cite{liu2017porous,lima2025th}, 2D Dewar-Anthracene demonstrates superior thermal resilience. This behavior is attributed to its fully conjugated carbon backbone and stable ring motifs, both of which contribute to the rigidity and resilience of the lattice, even in the presence of thermal agitation.

\begin{figure*}[!htb]
    \centering
    \includegraphics[width=\linewidth]{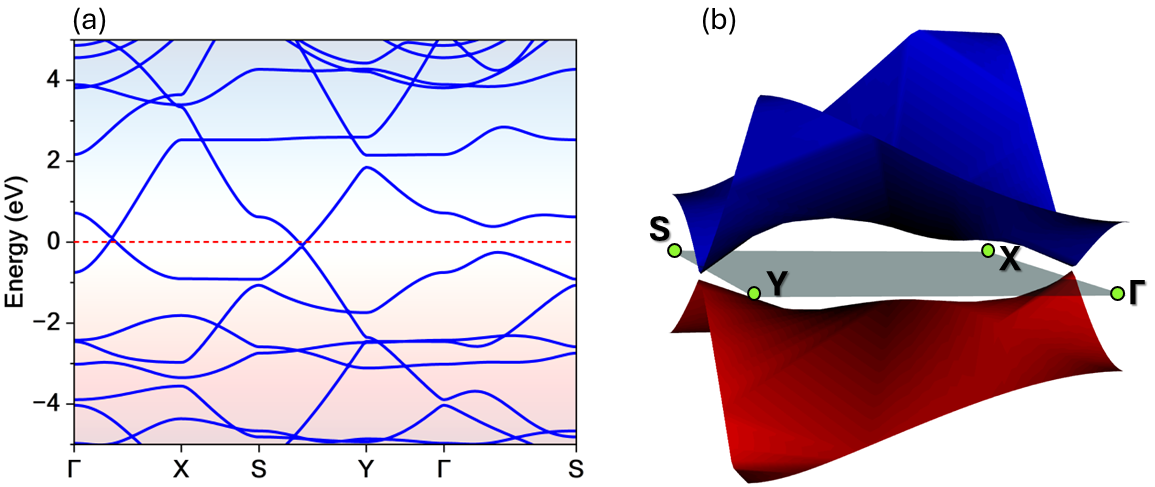}
    \caption{(a) 2DDA electronic band structure. (b) 3D dispersion of the valence and conduction bands near the $\Gamma$-point, indicating a Dirac-like crossing.}
    \label{fig:bands}
\end{figure*}

The 2DDA electronic properties reveal a metallic behavior driven by the extended $\pi$ states. As shown in Figure \ref{fig:bands}(a), two bands cross the Fermi level, forming Dirac-like crossings along the $\Gamma \rightarrow X$ and $S \rightarrow Y$ directions. These characteristics allow us to classify 2DDA as a Dirac metal, where charge carriers are expected to behave as massless Dirac fermions, leading to high electronic mobility. 

A remarkable feature of the electronic band structure is the presence of multiple Dirac-like points not only near the Fermi level but also in both the conduction and valence bands. These additional Dirac-like features suggest the existence of multiple high-mobility transport channels, which could enhance electronic performance in potential applications. Furthermore, the coexistence of dispersive and nearly flat bands near the Fermi level hints at possible electron correlation effects, which may play a role in unconventional transport phenomena.

Figure \ref{fig:bands}(b) illustrates the three-dimensional dispersion features around the Fermi level, providing deeper insights into the band structure characteristics. A well-defined cone-shaped intersection is observed, indicating the presence of massless Dirac fermions. Despite a slight tilt and direction-dependent anisotropy, the system retains the main features of the linear band dispersion near the Fermi level, which is a characteristic signature of Dirac-like materials. 

Such band crossings are particularly desirable for applications requiring high electron mobility, robust quantum transport, and anisotropic conductivity \cite{Trescher2015}. Similar electronic behavior has been reported for other carbon allotropes, such as phagraphene \cite{Wang2015Phagraphene} and T-graphene \cite{Liu2012TGraphene}. However, unlike these materials, 2DDA uniquely combines intrinsic porosity with remarkable thermodynamic stability. This combination enhances its potential use for selective molecular sieving and adsorption applications.

\begin{figure}[!htb]
    \centering
    \includegraphics[width=\linewidth]{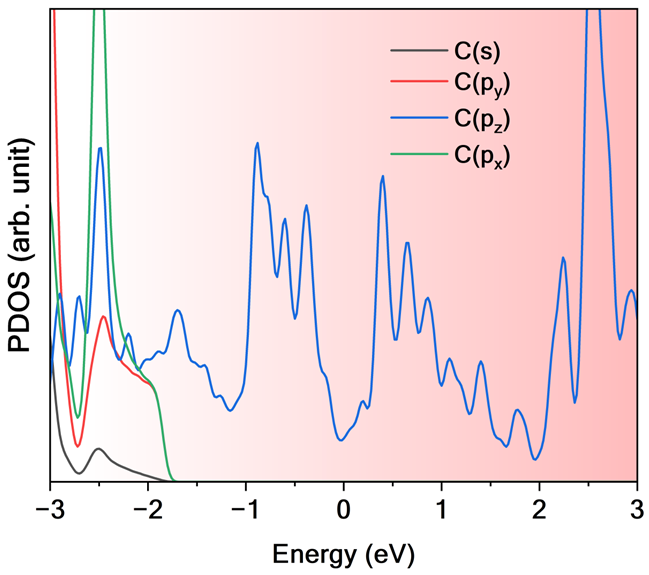}
    \caption{2DDA Projected density of states (PDOS) showing dominant $p_z$ contributions around the Fermi level. This PDOS figure plot refers to Figure \ref{fig:bands}.}
    \label{fig:pdos01}
\end{figure}

A closer look at the projected density of states (PDOS) clarifies the orbital nature of the 2DDA electronic states around the Fermi level. As shown in Figure \ref{fig:pdos01}, the electronic structure is dominated by the carbon $p_z$ orbitals (blue curve), which span the entire range near and across the Fermi level, confirming that the 2DDA conduction is primarily due to $\pi$-electron delocalization, a characteristic feature of sp$^{2}$-hybridized carbon systems.

In contrast, the $p_x$ and $p_y$ orbitals (green and red curves, respectively) contribute mainly below -2 eV and are less significant near the Fermi level, indicating no significant contributions to the conduction processes. Similarly, the $s$ orbitals (black curve) are localized deeper in the valence regions and contribute negligibly to the electronic transitions.

This orbital-resolved analysis supports the Dirac-like behavior observed in the electronic band structure (Figure \ref{fig:bands}), where the linear crossing at the $\Gamma \rightarrow X$ and $S \rightarrow Y$ directions emerges from $\pi$-dominated bands with minimal $s$/$\sigma$ mixing. The dominance of $p_z$ states also reinforces the planarity and extensive conjugation of the structure.

Compared to other porous 2D carbon materials like graphyne \cite{Narita1998Graphyne} or biphenylene network \cite{Tyutyulkov1997} (also known as Net-C \cite{Wang2013Metallic2D}), which often exhibit hybridization between $\pi$ and $\sigma$ states near the Fermi level due to acetylenic linkages or non-planar features, 2DDA remains distinctly $\pi$-dominated, with sharper and more isolated features in the PDOS.

\begin{figure}[!htb]
    \centering
    \includegraphics[width=\linewidth]{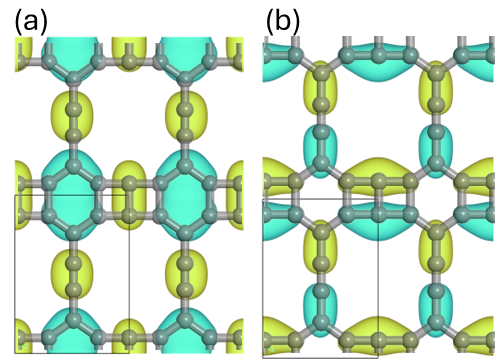}
    \caption{(a) 2DDA highest occupied crystalline orbital (HOCO) and (b) lowest unoccupied crystalline orbital (LUCO) one, illustrating the extensive delocalization of the frontier states.}
    \label{fig:orbitals}
\end{figure}

Visual representations of frontier electronic states are provided in Figure \ref{fig:orbitals}, with panel (a) showing the highest occupied crystalline orbital (HOCO) and panel (b) depicting the lowest unoccupied crystalline orbital (LUCO). These states dominate the reactivity and electronic transport, particularly near the Fermi level.

In the HOCO, the charge density is mainly concentrated along the dewar-anthracene-like chains, forming a continuous $\pi$-network in one crystallographic direction. This configuration suggests efficient in-plane hole mobility, with the potential for highly directional charge transport. On the other hand, LUCO extends over both the chains and the Dewar-type connectors, revealing a broader spatial distribution that could enable more isotropic or multidirectional electron conduction.

Such complementary patterns between the occupied and unoccupied frontier orbitals reveal an intrinsic electronic anisotropy, which resonates with the tilted Dirac cone observed in the electronic band structure (Figure \ref{fig:bands}). This anisotropy is especially advantageous for direction-selective device applications, including field-effect transistors and anisotropic optoelectronic platforms.

\begin{figure}[!htb]
    \centering
    \includegraphics[width=0.7\linewidth]{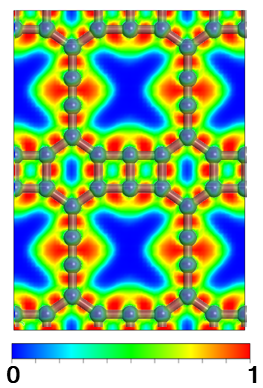}
    \caption{2DDA electron localization function (ELF). Large ELF values indicate strong $\sigma$-bonds, while low ones reveal extended $\pi$-delocalization.}
    \label{fig:elf}
\end{figure}

Electron pairing and bonding characteristics within 2DDA can be understood through the electron localization function (ELF), illustrated in Figure \ref{fig:elf}. Regions of large ELF values (in red) represent areas of strong electron localization, typically associated with $\sigma$-bonding, while low values (blue) correspond to delocalized electrons contributing to $\pi$-bonding.

A clear pattern emerges in which the C-C $\sigma$-bonds inside the anthracene units exhibit strong localization, ensuring mechanical integrity along the chains. In contrast, the linking Dewar-type rings and the spaces among the anthracene units display broader, more diffuse ELF values, a characteristic of $\pi$-electron delocalization throughout the 2D lattice and the torsional tension associated with these motifs. This contrast reveals a hybrid bonding environment: locally confined $\sigma$-bonds embedded within a globally delocalized $\pi$-network. Such duality is a common feature in conjugated carbon systems like graphene \cite{CastroNeto2009GrapheneReview}.

\begin{figure}[!htb]
    \centering
    \includegraphics[width=\linewidth]{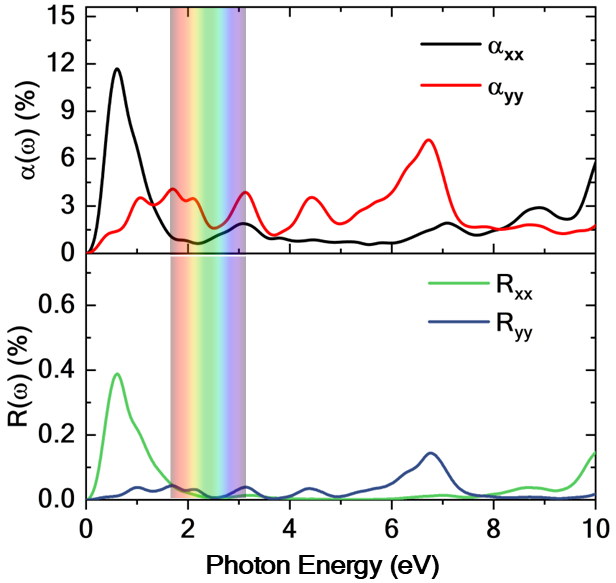}
    \caption{2DDA direction-resolved optical absorption (top) and reflectivity (bottom) spectra under $x$- and $y$-polarized lights.}
    \label{fig:opticalprop}
\end{figure}

Strong direction-dependent optical behavior is evident in the absorption spectra, as shown in the top panel of Figure \ref{fig:opticalprop}. The imaginary part of the dielectric function, represented here as the absorption coefficient $\alpha$, reveals prominent peaks for both $x$- and $y$-polarized light. The $\alpha_{xx}$ component reaches values above 12\% near 1.6 eV, while the $\alpha_{yy}$ curve exhibits a broader, multi-peaked response throughout the visible and ultraviolet ranges. These trends confirm that 2DDA is optically active and exhibits significant optical anisotropy.

The presence of well-defined absorption features in the 1.5–3.5 eV range —overlapping the visible spectrum shaded in the figure — suggests potential applications in photodetectors, light-harvesting systems, and transparent conducting films. The anisotropic nature of these peaks could allow for polarization-sensitive optoelectronics, where the response is tuned to incident light direction.

On the bottom panel of Figure \ref{fig:opticalprop}, the reflectivity spectra $R$ show low overall values, below 0.6\%, indicating that 2DDA is highly transmissive in the investigated range. This transparency is especially relevant for coating or filtering applications, where minimal light loss is required. Reflectivity is slightly stronger along the $x$-direction near 1.8 eV, reinforcing the trend of optical asymmetry.

Altogether, these optical responses highlight a balance between light absorption and transmission, with direction-specific tunability. This behavior stems from the anisotropic electronic structure and delocalized $\pi$-network discussed previously, which governs both the excitation dynamics and dielectric response.

\begin{figure}[!htb]
    \centering
    \includegraphics[width=\linewidth]{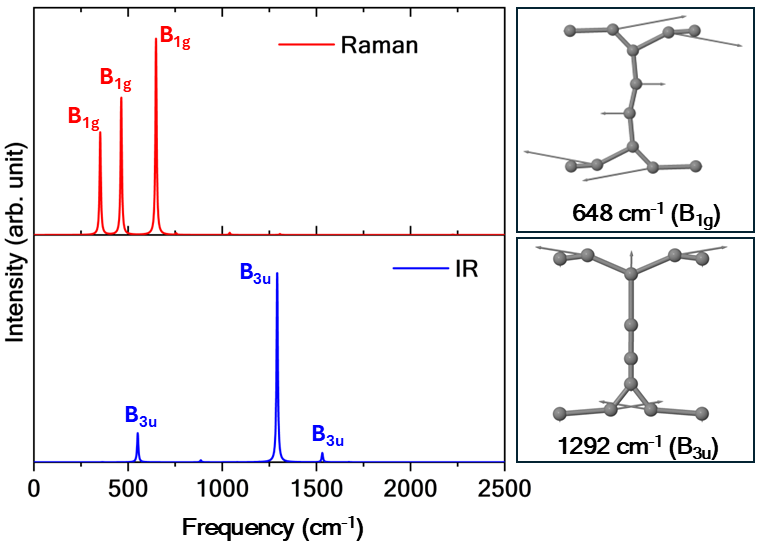}
    \caption{2DDA simulated Raman (top) and infrared (bottom) spectra with labeled vibrational modes.}
    \label{fig:Raman+IR}
\end{figure}

A 2DDA detailed vibrational fingerprint is provided in Figure \ref{fig:Raman+IR}. The 2DDA vibration modes are given by $\Gamma_{vib}$ = 5$A_{g}$ + 5$B_{1g}$ + 4$B_{2g}$ + $B_{3g}$ +$A_{u}$ + 4$B_{1u}$ + 5$B_{2u}$ + 5$B_{3u}$. Three modes are acoustic ($B_{1u}$ + $B_{2u}$ + $B_{3u}$), 15 are Raman-active (5$A_{g}$ + 5$B_{1g}$ + 4$B_{2g}$ + $B_{3g}$), and 14 are infrared-active (4$B_{1u}$ + 5$B_{2u}$ + 5$B_{3u}$). 

The Raman spectrum (top panel) displays sharp and well-separated peaks, with dominant modes labeled by their corresponding symmetries. The three most intense bands can be identified in the spectrum at 352 cm$^{-1}$ ($B_{1g}$), 463 cm$^{-1}$ ($B_{1g}$), and 648 cm$^{-1}$ ($B_{1g}$). The vibration associated with the band at 648 cm$^{-1}$ is denoted by an asymmetric being at the acetylenic group and by the asymmetric stretching on the benzene units in 2DDA. Lower-frequency Raman peaks are characteristic of extended aromatic frameworks and resemble features observed in graphene derivatives and polycyclic hydrocarbons \cite{Ferrari2013Raman,Cloutis2016PAH}.

In the infrared (IR) spectrum shown in the lower panel of Figure \ref{fig:Raman+IR}, several active modes are observed, most notably intense peaks near 551 cm$^{-1}$ ($B_{3u}$), 1292 cm$^{-1}$ ($B_{3u}$), and 1531 cm$^{-1}$ ($B_{3u}$). As illustrated by the vibration at 1531 cm$^{-1}$, these modes are related to symmetric and asymmetric stretching along the C-C and C=C bonds. 

%Analyzing the Raman spectrum (Fig. \ref{fig:Raman+IR}, top panel) reveals several distinct peaks, with the most intense modes $Ag$ and $B_{1g}$ occurring at low and intermediate frequencies. Four prominent bands are visible at 985 cm$^{-1}$ ($A_{g}$), 1172 cm$^{-1}$ ($A_{g}$), 1383 cm$^{-1}$ ($B_{1g}$), and 2253 cm$^{-1}$ ($A_{g}$). Strong $A_g$ symmetric modes suggest collective in-plane stretching and breathing vibrations of the carbon framework. The sharp intensity of these peaks highlights the high rigidity of the system, which is commonly observed in sp$^{2}$-hybridized carbon networks. The prominent peak at the high-frequency region corresponds to a strong C--C stretching mode, a signature of a well-preserved conjugated structure in the material.

%A detailed vibrational fingerprint of 2D Dewar-Anthracene is provided in Figure \ref{fig:Raman+IR}. The Raman spectrum (top panel) displays sharp and well-separated peaks, with dominant modes labeled by their corresponding symmetries. A strong peak around 2100 cm$^{-1}$, attributed to an A$_g$ symmetric stretching mode, dominates the spectrum, suggesting a highly polarizable vibration along the conjugated backbone of the structure.

The coexistence of IR- and Raman-active modes with well-defined frequencies reflects the high symmetry of the 2D lattice. At the same time, the presence of non-overlapping peaks in the two spectra confirms the mutual exclusion rules expected in centrosymmetric structures. Importantly, these sharp vibrational fingerprints provide a unique spectral signature that would facilitate the 2DDA experimental detection through Raman and IR spectroscopy.

\begin{figure*}[!htb]
    \centering
    \includegraphics[width=\linewidth]{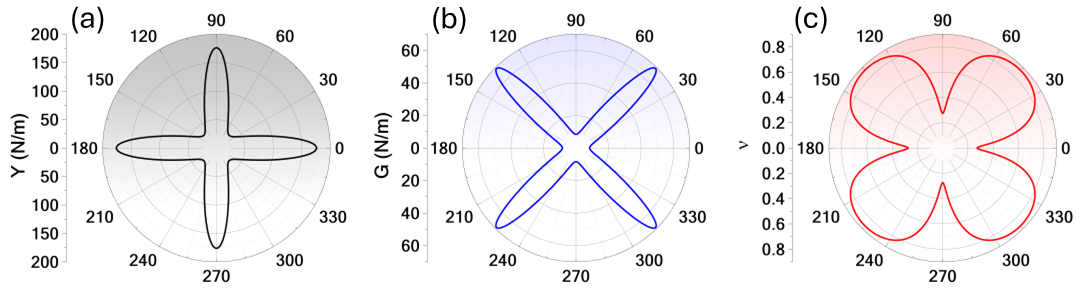}
    \caption{2DDA polar plots of (a) Young’s modulus ($Y$), (b) Shear modulus ($G$), and (c) Poisson's ratio ($\nu$). }
    \label{fig:mechprop}
\end{figure*}

2DDA mechanical properties anisotropy can be quantitatively verified from the polar plots of Figure \ref{fig:mechprop}. Figure \ref{fig:mechprop}(a) shows the directional dependence of Young's modulus (Y). The material exhibits a clear four-lobed pattern, with stiffness peaks along the 0$^\circ$ and 90$^\circ$ directions and minima along 45$^\circ$, indicating a strong orthotropic mechanical behavior.

The highest stiffness direction, described by the Young modulus ($Y$), reaches approximately 176.24 N/m, while the perpendicular direction shows a slightly lower value of 31.51 N/m. A high anisotropy is verified, with a reason of 5.59, denoting strong directional dependence of $Y$ These values are in the same range of other porous carbon allotropes, such as penta-graphene (263.8 N/m) \cite{Li2016PentaGraphene}, phagraphene (150-260 N/m) \cite{Pereira2016PhagrapheneMD,Sun2016PentaPhagraphene,Wu2017PhagrapheneModulation}, graphyne (162 N/m) \cite{Peng2012Graphyne}, graphdiyne (150-170 N/m) \cite{Ahangari2015Graphdiyne,Polyakova2024Elastic}, despite 2DDA having a more open porous framework. 

In Figure \ref{fig:mechprop}(b), the shear modulus (G) also displays a strong anisotropy with a characteristic “X” shape. Maxima (69.14 N/m) and minima (8.43 N/m) alternate every 45$^{\circ}$, reflecting variations in angular stiffness related to the 2DDA topology and acetylene motifs in the lattice. This behavior is especially promising for designing direction-sensitive mechanical components or strain-tunable devices.

The Poisson ratio $\nu$, plotted in Figure \ref{fig:mechprop}(c), shows a butterfly-like four-lobed distribution, ranging from 0.27 to 0.87, depending on the direction of applied stress and transverse response. Values close to 0.87 are unusually high for carbon-based 2D materials and suggest high transverse compliance along specific orientations \cite{Jana2021BeyondGraphene}.

\subsection{2DDA Nanoribbons}

\begin{figure*}[!htb]
    \centering
    \includegraphics[width=\linewidth]{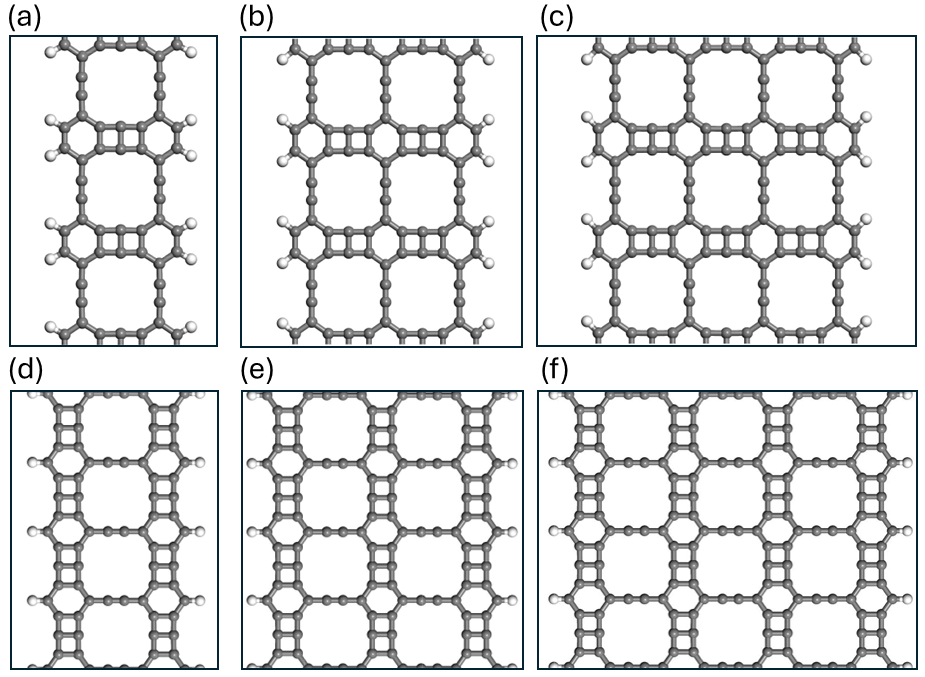}
    \caption{Atomic structures of hydrogen-passivated nanoribbons derived from 2DDA. (a–c) diphenylacetylene edge type; (d–f) dewar-anthracene edge type. The width increases from left to right by adding one rectangular pore unit.}
    \label{fig:nanoribbons}
\end{figure*}

As for some carbon allotropes, the nanoribbons synthesis proved to be easier than large layers, we have also investigated 2DDA finite fragments (nanoribbons). In Figure \ref{fig:nanoribbons}, we present six nanoribbons of different widths (but infinite along y-directions) derived from 2DDA and grouped according to their edge topology and increasing width. 

Figures \ref{fig:nanoribbons}(a)–(c) represent nanoribbons terminated by what we call "diphenylacetylene edges," characterized by the presence of a six-membered ring at the edge, followed by a short linear chain of carbon atoms. This edge type resembles the termination commonly seen in acene-like systems, where extended $\pi$-conjugation is preserved along the ribbon direction \cite{Narita2015Nanographene}. As the ribbon width increases from (a) to (c), one rectangular pore is added symmetrically, progressively recovering the periodicity of the 2D lattice.

Conversely, Figures \ref{fig:nanoribbons}(d)–(f) show nanoribbons with what we call  "dewar-anthracene edges", comprising two adjacent four-membered rings and a six-membered ring near the ribbon boundary. This edge motif introduces more geometric frustration and curvature due to the presence of smaller rings, which can influence both the structural relaxation and the electronic localization near the edge. Again, the width increases with the addition of rectangular pore units, maintaining topological consistency across the series.

\begin{figure*}[!htb]
    \centering
    \includegraphics[width=\linewidth]{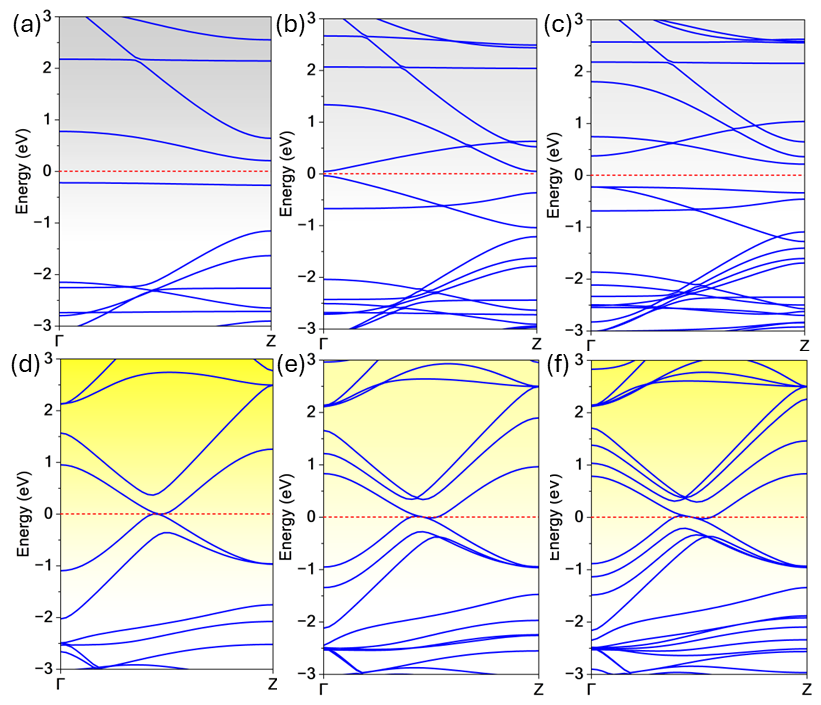}
    \caption{Electronic band structures of 2DDA nanoribbons of increased width, from left to right. (a–c): diphenylacetylene-type edges; (d–f): dewar-anthracene-type.}
    \label{fig:bands1d}
\end{figure*}

Figure \ref{fig:bands1d} presents the electronic band structures of six nanoribbons derived from 2D Dewar-Anthracyne, highlighting the effects of edge geometry and width scaling on their electronic behavior. Panels (a–c) correspond to nanoribbons with diphenylacetylene-type edges, while (d–f) correspond to dewar-anthracene-type ones.

The electronic structure of the narrowest diphenylacetylene-type nanoribbon (Figure \ref{fig:bands1d}(a)) exhibits a semiconducting character with an indirect band gap of 0.42 eV, which can be attributed to strong quantum confinement and reduced conjugation of the narrow width. For the nanoribbon presented in Figure \ref{fig:bands1d}(b), the bandgap decreases, eventually reaching a quasi-metallic character, with conduction and valence bands approaching each other near the Fermi level. In Figure \ref{fig:bands1d}(c), the electronic bandgap is  0.42 eV. This width-dependent electronic transition resembles what is observed in armchair graphene nanoribbons \cite{son2006energy}. The dewar-anthracene-type nanoribbons (Figures \ref{fig:bands1d}d–f) display markedly different behavior. They all show metallic features, with bands crossing at the Fermi level, reinforcing the idea that dewar-anthracene-type nanoribbons behave like zigzag graphene nanoribbons \cite{son2006energy}.

\begin{figure*}[!htb]
    \centering
    \includegraphics[width=\linewidth]{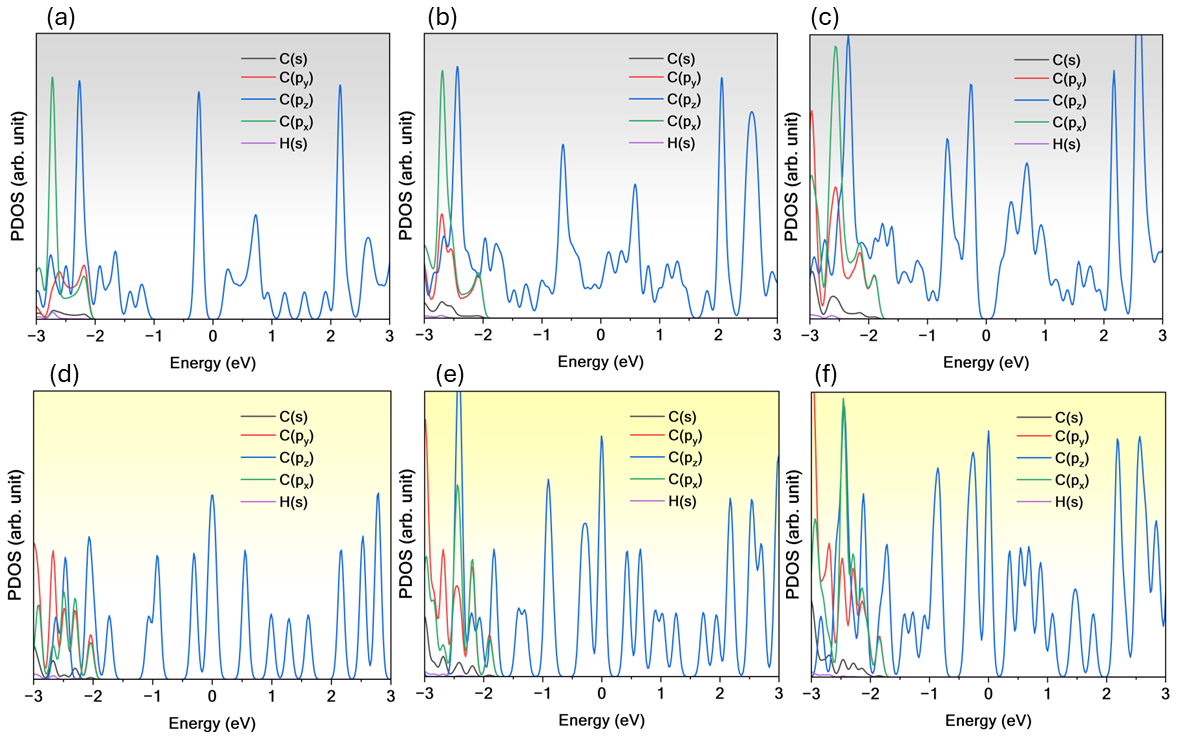}
    \caption{Projected density of states of 2DDA nanoribbons. (a–c): diphenylacetylene-type edges; (d–f): dewar-anthracene-type edges. }
    \label{fig:pdos02}
\end{figure*}

In Figure \ref{fig:pdos02}, we present the corresponding PDOS for the six nanoribbons shown in Figure \ref{fig:bands1d}. These results clarify the orbital contributions near the Fermi level and reveal how quantum confinement, edge geometry, and ribbon width influence the electronic features.

For the diphenylacetylene-type nanoribbons (panels a-c), the PDOS highlights the dominant contributions from carbon $p_z$ orbitals around the Fermi level, confirming that $\pi$-electron conjugation is preserved and dominates the frontier states. Figure \ref{fig:pdos02}(a) shows a clear energy bandgap with no states near the Fermi level, consistent with the semiconducting behavior observed in the band structure. As width increases (Figures \ref{fig:pdos02}(b,c)), the $p_z$ states begin to fill the gap region, indicating increased electronic delocalization. In Figure \ref{fig:pdos02}(c), the presence of finite PDOS at the Fermi level suggests the system transitions to a narrow-gap semiconductor.

The dewar-anthracene-type nanoribbons (Figures \ref{fig:pdos02} (d–f)) show metallic PDOS profiles across all widths. In each case, $p_z$ states contribute directly at the Fermi level, with high density and continuous features. This confirms that the delocalized edge states from the Dewar-type motifs persist even at narrow ribbon widths. The $p_x$ and $p_y$ orbitals remain largely inactive near the Fermi level, consistent with their limited role in out-of-plane conjugation.

\section{Conclusion}

In summary, a new 2D carbon allotrope was proposed, combining the topology of Dewar-anthracene and graphynes, named 2D Dewar-Anthracyne (2DDA). 2DDA consists of chains of Dewar-anthracenes connected by acetylenic linkages. DFT-based simulations confirmed its structural stability with a cohesive energy of -7.13 eV/atom and its dynamic by the absence of imaginary phonon modes. Also, Ab-initio molecular dynamics simulations at 300 K further confirmed its thermal robustness.

The metallic character of 2DDA was demonstrated by a Dirac-like cone at the $\Gamma$-point and a PDOS dominated by $2p_z$ orbitals, presenting extended $\pi$-conjugation. Young’s modulus and shear modulus range from 31.51 to 176.24 N/m and from 8.43 to 69.14 N/m, respectively. Also the Poisson ratio goes from 0.27 to 0.87, indicating significant mechanical anisotropy. The optical absorption spectra show activity in the visible and UV regions, with low reflectivity ($< 0.6$\%) and direction-dependent response. Simulated Raman and IR spectra exhibit sharp and well-separated peaks, providing a clear vibrational fingerprint.

Nanoribbons derived from 2DDA exhibit distinct electronic behavior depending on edge termination type. Diphenylacetylene-type nanoribbons display indirect and direct band gaps ranging from 0.60 to 0.40 eV, while dewar-anthracene-type ones are consistently metallic. These combined properties highlight the 2DDA potential for applications in flexible nanoelectronics and optoelectronic devices.

\section*{Data access statement}
Data supporting the results can be accessed by contacting the corresponding author.

\section*{Conflicts of interest}
The authors declare no conflict of interest.

\section*{Acknowledgements}
This work was supported by the Brazilian funding agencies Fundação de Amparo à Pesquisa do Estado de São Paulo (FAPESP) (grants no. 2022/03959-6, 2022/14576-0, 2013/08293-7, 2020/01144-0, 2024/05087-1, and 2022/16509-9), National Council for Scientific, Technological Development (CNPq) (grants no. 307213/2021–8, 350176/2022-1, and 167745/2023-9), FAP-DF (grants no. 00193.00001808/2022-71 and 00193-00001857/2023-95), FAPDF-PRONEM (grant no. 00193.00001247/2021-20), and PDPG-FAPDF-CAPES Centro-Oeste (grant no. 00193-00000867/2024-94). The authors acknowledge the Molecular Simulation Laboratory at São Paulo State University (UNESP) and the Center for Computing in Engineering and Sciences at Unicamp.

\printcredits

\bibliography{biblio/cas-refs}

\end{document}